\documentclass{synmet}

\usepackage{epsfig}

\usepackage{amssymb}

 \advance\topmargin by -18mm

\begin{document}

\begin{frontmatter}

% Title, authors and addresses

% use the thanksref command within \title, \author or \address for footnotes;
% use the corauthref command within \author for corresponding author footnotes;
% use the ead command for the email address,
% and the form \ead[url] for the home page:
% \title{Title\thanksref{label1}}
% \thanks[label1]{}
% \author{Name\corauthref{cor1}\thanksref{label2}}
% \ead{email address}
% \ead[url]{home page}
% \thanks[label2]{}
% \corauth[cor1]{}
% \address{Address\thanksref{label3}}
% \thanks[label3]{}

\title{\bf How to detect edge electron states in (TMTSF)$\bf_2$X and
$\bf Sr_2RuO_4$ experimentally}

% use optional labels to link authors explicitly to addresses:
% \author[label1,label2]{}
% \address[label1]{}
% \address[label2]{}

\author{Hyok-Jon~Kwon,}
\author{Victor M.~Yakovenko,}
\author{and K. Sengupta\thanksref{label1}} 

\thanks[label1]{Present address: Department of Physics, Yale
University,
New Haven, CT 06520-8120.}

\address{Department of Physics and Center for
  Superconductivity Research, University of Maryland, College Park, MD
  20742, USA}

\begin{abstract}
We discuss a number of experiments that could detect the electron edge
states in the organic quasi-one-dimensional conductors $\rm(TMTSF)_2X$
and the inorganic quasi-two-dimensional perovskites $\rm Sr_2RuO_4$.
We consider the chiral edges states in the magnetic-field-induced
spin-density-wave (FISDW) phase of $\rm(TMTSF)_2X$ and in the
time-reversal-symmetry-breaking triplet superconducting phase of $\rm
Sr_2RuO_4$, as well as the nonchiral midgap edge states in the triplet
superconducting phase of $\rm(TMTSF)_2X$.  The most realistic
experiment appears to be an observation of spontaneous magnetic flux
at the edges of $\rm Sr_2RuO_4$ by a scanning SQUID microscope.
\end{abstract}

\begin{keyword}
% keywords here, in the form: keyword \sep keyword
  Many-body and quasiparticle theories \sep
  Transport measurements, conductivity, Hall effect, magnetotransport \sep
  Organic conductors based on radical cation and/or anion salts \sep
  Organic superconductors \sep
  Ruthenate superconductors
% PACS codes here, in the form: \PACS code \sep code
\PACS 74.70.Kn \sep % Organic superconductors
74.70.Pq \sep       % Ruthenates 
73.43.-f \sep       % Quantum Hall effects
73.20.-r            % Electron states at surfaces and interfaces
\hfill\bf cond-mat/0111071, v.1: 5 November 2001, v.2: 15 November 2001
%\hfill\bf Version T.2, edited by VMY 11/15/2001, compiled \today 
\end{keyword}
\end{frontmatter}

\section{Introduction}

Edge electron states in various materials attracted a great deal of
attention recently.  In this paper, we discuss some experiments
proposed to observe the effects of such states in the organic
quasi-one-dimensional (Q1D) conductors $\rm(TMTSF)_2X$ and the
inorganic quasi-two-dimensional perovskites $\rm Sr_2RuO_4$.  For
general overviews of these materials, see Refs.\ \cite{TMTSF} and
\cite{PhysToday}, respectively.  The purpose of the paper is to
encourage practical realization of these experiments by summarizing
basic ideas and giving quantitative order-of-magnitude estimates
without going into deep theoretical physics and mathematical
formalism.  The latter can be found in the cited references.

In general, a system with an energy gap has delocalized electron
states in the bulk with energies above and below the energy gap.
However, it may also have bound states with energies inside the gap,
which are localized near the sample edges or other inhomogeneities.
The energy gap may be of different origin, e.g.\ insulating or
superconducting.  In Sec.\ \ref{sec:FISDW}, we study the case where
the gap is produced by the magnetic-field-induced spin-density wave
(FISDW) in $\rm(TMTSF)_2X$.  In Secs.\ \ref{sec:midgap} and
\ref{sec:SrRu}, we consider triplet superconducting states in
$\rm(TMTSF)_2X$ and $\rm Sr_2RuO_4$.

%%%%%%%%%%%%%%%%%%%%%%%%%%%%%%%%%%%%%%%%%%%%%%%%%%%%%%%%%%%%%%%%%%%%%
\section{Chiral edge states in the FISDW phase of $\rm(TMTSF)_2X$ }
\label{sec:FISDW}
%%%%%%%%%%%%%%%%%%%%%%%%%%%%%%%%%%%%%%%%%%%%%%%%%%%%%%%%%%%%%%%%%%%%%

$\rm(TMTSF)_2X$ are Q1D crystals consisting of conducting chains
parallel to the {\bf a} axis, with substantial interchain coupling in
the {\bf b} direction and much weaker coupling in the {\bf c}
direction \cite{TMTSF}, along which we select the $x$, $y$, and $z$
axes.  The lattice spacings are $a=0.73$ nm, $b=0.77$ nm, and $c=1.35$
nm, whereas the typical sample dimensions are $L_x=2$ mm and
$L_y\approx L_z=0.2$ mm.  Thus, a typical sample contains
$\mathcal{N}_c=L_yL_z/bc\approx 4\times10^{10}$ chains and
$\mathcal{N}_l=L_z/c\approx1.5\times10^5$ layers.  When a magnetic
field $H$ is applied along the {\bf c} axis, it causes a phase
transition into the FISDW state, which exhibits the integer quantum
Hall effect (IQHE).  The Hall conductivity per one ({\bf a},{\bf b})
layer is $\sigma_{xy}=2Ne^2/h$, where $N$ is a small, $H$-dependent
integer number, $e$ is the electron charge, and $h$ is the Planck
constant.  In general, it is expected that $N$ chiral gapless edge
states should exist in an IQHE system along the perimeter of the
sample, as sketched in Fig.\ \ref{fig:FISDW}.  Such edge states were
indeed constructed theoretically for FISDW in Refs.\ \cite{Goan,tof}.

\subsection{Time-of-flight experiment}
\label{pulse1}

\begin{figure}[t]
\centerline{\psfig{file=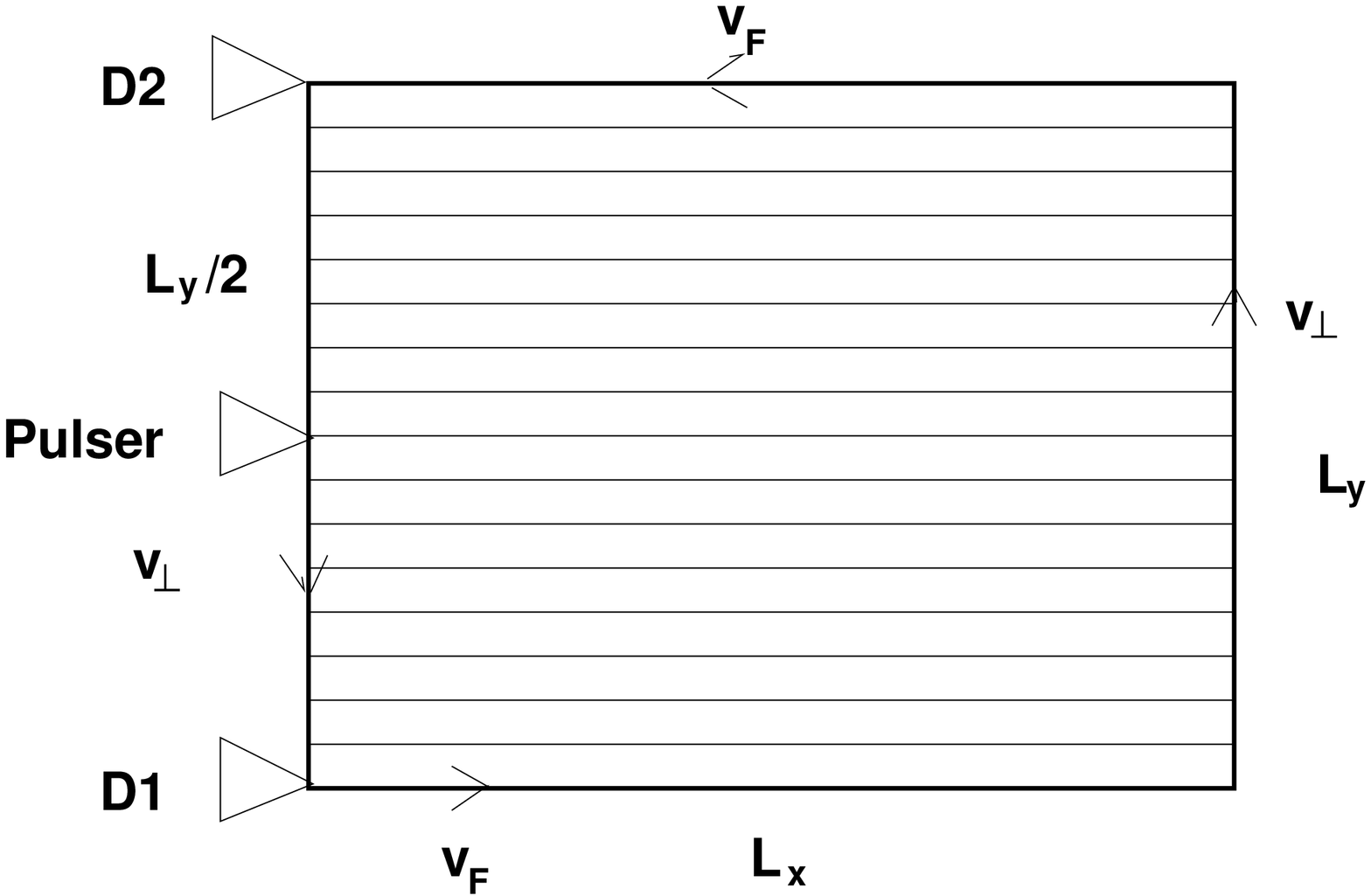,width=0.7\linewidth}}
\caption{Sketch of the proposed time-of-flight experiment. The arrows
indicate the directions of the edge states velocities $v_{\perp}$ and
$v_F$.  The thin lines indicate the conducting chains of
$\rm(TMTSF)_2X$.  The pulser sends a pulse, which is detected at
different times by the detectors D1 and D2.}
\label{fig:FISDW}
\end{figure}

As was shown in Refs.\ \cite{Goan,tof}, the edge states travel with
the group velocities $v_\perp=Nb\Delta/\hbar\approx N\times300$ m/s
perpendicular to the chains and the Fermi velocity $v_F\approx190$
km/s parallel to the chains, where $\Delta\approx 3$ K is the FISDW
gap.  In the time-of-flight experiment described in Ref.\ \cite{tof},
a pulse, sent by the pulser, perturbs the edge states (see Fig.\
\ref{fig:FISDW}).  This perturbation is carried downstream and reaches
the detector D1 with the delay time $t=L_y/2v_\perp\approx0.33$ $\mu$s
and the detector D2 with the greater delay $3t$.  The pulse can be
electric (perturbing the occupation number of the edge states), or
magnetic (creating spin polarization of the edge states), or thermal
(perturbing the electron temperature of the edge states).  The
time-of-flight experiments with electric perturbations have been
successfully performed in semiconducting QHE systems \cite{Ashoori}.

\subsection{Specific heat}

Since the chiral edge states are gapless, their the specific heat
$C_e$ is linear in temperature $T$ and should dominate at low
temperatures, where the bulk contribution is frozen out because of the
FISDW gap $\Delta$.  The edge specific heat per one chain is
$C_e/T=2\pi k_B^2/3\Delta\approx10^{-23}$ $\rm J/K^2$, where $k_B$ is
the Boltzmann constant \cite{tof}.  Multiplying this number by the
total number of chains $\mathcal{N}_c\approx 4\times10^{10}$, we
obtain the total edge specific heat $4\times10^{-13}$ $\rm J/K^2$.
This value is smaller than the bulk specific heat in the normal state
by the factor $2\xi/L_x\approx0.5\times10^{-3}$ (where $\xi=\hbar
v_F/\Delta\approx0.5$ $\mu$m is the coherence length), but it may be
still measurable with a sensitive technique \cite{Scheven95}.

\subsection{Thermal quantum Hall effect}
\label{ThermalQH1}

\begin{figure}[t]
\centerline{\psfig{file=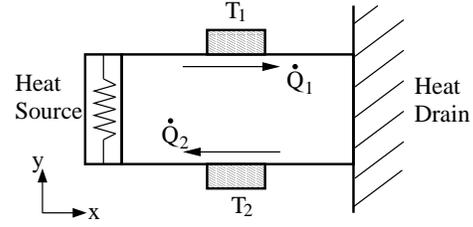,width=0.8\linewidth}}
\caption{Sketch of the thermal quantum Hall effect experiment.  The
heat currents $\dot{Q}_1$ and $\dot{Q}_2$ are carried along the edges
with the temperatures $T_1$ and $T_2$.}
\label{fig:TQHE}
\end{figure}

1D chiral electron gas of temperature $T$ carries thermal current
$\dot{Q}=\pi^2k_B^2T^2/6h$.  Let us consider a FISDW sample with a
heat source on the left, a heat drain on the right, and two
thermometers $T_1$ and $T_2$ on the sides, as show in Fig.\
\ref{fig:TQHE}.  The circulating edge current raises its temperature
to $T_1$ by gaining thermal energy from the heat source and flows
along the top edge of the bar maintaining that temperature (assuming
no heat loss) until it reaches the heat drain.  There the edge current
loses its thermal energy, drops its temperature to $T_2$, and returns
along the bottom edge maintaining that temperature.  The net heat
current from the source to the drain is
\begin{equation}
  \dot{Q}_x=\dot{Q}_1-\dot{Q}_2 =K_{xy}\,\delta_yT,\;\;\;
  \frac{K_{xy}}{T}=2N\frac{\pi^2k_B^2}{3h},
  \label{Hall}
\end{equation}
where $K_{xy}$ is the thermal Hall conductance, $\delta_y T=T_1-T_2$
is the temperature difference in the $y$ direction, and
$T=(T_1+T_2)/2$ is the average temperature of the edges.  Eq.\
(\ref{Hall}) demonstrates the thermal QHE proposed in Ref.\
\cite{Kane}.  Both the thermal and electrical Hall conductances are
quantized with the same integer $N$ and are related to each other by
the Wiedemann-Franz law for free electrons \cite{Kane}.  Detection of
the thermal QHE will therefore confirm the existence of the chiral
edge states in the FISDW phase. The quantum of thermal Hall
conductance per one layer is $\pi^2k_B^2/3h=0.946\times10^{-12}$ $\rm
W/K^2$. The total thermal Hall conductance is obtained by multiplying
this number by the number of layers
$\mathcal{N}_l\approx1.5\times10^5$.  The quantum of thermal
conductance has been measured in the state-of-the-art experiments
\cite{Schwab01}.

The above consideration assumed an idealized case where the
longitudinal thermal conductances $K_{xx}$ and $K_{yy}$ are zero. In
general, there will be nonzero temperature gradients in both $x$ and
$y$ directions:
\begin{equation}
  (\delta_xT,\delta_yT)={\dot{Q}_x \over K_{xx}K_{yy}+K_{xy}^2} 
  (K_{yy},K_{xy}).
  \label{Ktensor}
\end{equation}
Because of the Wiedemann-Franz law, we expect that $K_{xx}K_{yy}\ll
K_{xy}^2$ in the QHE regime, where
$\sigma_{xx}\sigma_{yy}\ll\sigma_{xy}^2$.  Then Eq.\ (\ref{Hall})
approximately holds.  However, there also exist phonon contributions
to $K_{xx}$ and $K_{yy}$. In (TMTSF)$_2$ClO$_4$, the longitudinal
thermal conductance per layer due to phonons is $ K_{xx}/T \approx
5\times 10^{-11}\rm ~W/K^2$ at $T\approx 0.2~\rm K$ and shows a
power-law temperature dependence \cite{Belin97}.  The phonon
contribution is comparable to the quantum of $K_{xy}$ and can be made
even smaller at lower temperatures, so an observation of the thermal
QHE in the FISDW state of $\rm(TMTSF)_2X$ is feasible.  In general,
the thermal Hall conductance can be determined from the relation
$K_{xy}=K_{yy}\,\delta_yT/\delta_xT$ \cite{Ong}.

%%%%%%%%%%%%%%%%%%%%%%%%%%%%%%%%%%%%%%%%%%%%%%%%%%%%%%%%%%%%%%%%%%%%%
\section{Midgap Andreev edge states in the triplet 
superconducting phase of $\rm(TMTSF)_2X$}
\label{sec:midgap}
%%%%%%%%%%%%%%%%%%%%%%%%%%%%%%%%%%%%%%%%%%%%%%%%%%%%%%%%%%%%%%%%%%%%%

\begin{figure}[t]
\centerline{\psfig{file=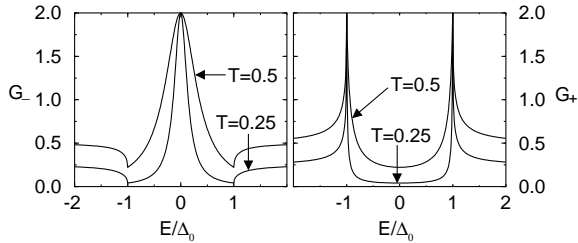,width=\linewidth}}
\caption{Conductances $G$ as functions of voltage $eV=E$ calculated
for tunneling into the edges perpendicular (left panel) and parallel
(right panel) to the chains in $\rm(TMTSF)_2X$.  Here $T$ is not a
temperature, but the transmission coefficient of tunneling barrier.}
\label{fig:tunnel}
\end{figure}

In $\rm(TMTSF)_2X$, the upper critical magnetic field $H_{c2}$ exceeds
the Pauli paramagnetic limit by a factor greater than 4 \cite{Lee97},
and the Knight shift does not change between the normal and
superconducting states \cite{Lee99}.  This indicates the triplet
character of superconductivity in these materials
\cite{theory,Lebed00}.  The triplet order parameter can be written as
$\langle\psi_\alpha({\bf k})\psi_\beta(-{\bf k})\rangle \propto
\epsilon_{\alpha\bar{\alpha}}
[\mbox{\boldmath$\sigma$}_\beta^{\bar{\alpha}}\cdot
\mbox{\boldmath$\Delta$}({\bf k})]$, where $\alpha$ and $\beta$ are
the spin indices, {\boldmath$\sigma$} are the spin Pauli matrices, and
$\epsilon_{\alpha\bar{\alpha}}=i\sigma^y_{\alpha\bar{\alpha}}$ is the
antisymmetric spin metric tensor.  We only consider the case of a
uniform spin orientation: $\mbox{\boldmath$\Delta$}({\bf k})={\bf
d}\,\Delta({\bf k})$, where {\bf d} is a unit vector.  The simplest
triplet pairing potential $\Delta({\bf k})$ is an odd function of
$k_x$, i.e.\ it has opposite signs on the two sheets of the Fermi
surface located near $\pm k_F$.  This sign change results in formation
of Andreev bound states with energies exactly in the middle of the
energy gap (midgap states) at the edges perpendicular to the chains,
as explained in Ref.\ \cite{Igor} (see also \cite{Tanuma}).  Unlike in
the FISDW case discussed in Sec.\ \ref{sec:FISDW}, these midgap edge
states are not chiral.

\subsection{Tunneling experiment}

Conceptually, the most straightforward way to detect the midgap edge
states is by electron tunneling between a normal metal and the
superconducting $\rm(TMTSF)_2X$.  Tunneling into the edges
perpendicular and parallel to the chains should exhibit a zero-bias
peak and a gap, as shown in the left and right panels of Fig.\
\ref{fig:tunnel}, correspondingly.  Unfortunately, it turned out
difficult to achieve good tunneling junctions with the organic
$\rm(TMTSF)_2X$.  Below we discuss alternative experiments.

\subsection{Paramagnetic spin response}

\begin{figure}[t]
\centerline{\psfig{file=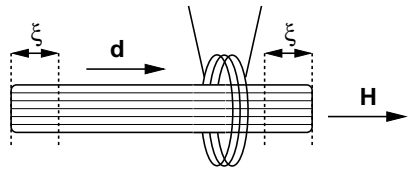,width=0.7\linewidth}}
\caption{Schematic experimental setup to measure magnetic
susceptibility of the edge states localized at the ends of the chains
in superconducting $\rm(TMTSF)_2X$.}
\label{fig:coil}
\end{figure}

When a magnetic field $H$ is applied parallel to the vector {\bf d},
the energies on the spin-up and spin-down midgap states split because
of the Zeeman effect.  At zero temperature, only the lower-energy
states would be occupied, thus the edge states should be
spin-polarized, yielding the magnetic moment
$\mu_B/2=4.6\times10^{-24}$ A~m$^2$ per chain ($\mu_B$ is the Bohr
magneton), or $\mathcal{N}_c\mu_B/2=1.8\times10^{-13}$ A~m$^2$ for the
whole edge \cite{Igor}.  At a finite temperature, the edges would
exhibit a Curie-like paramagnetic spin response to the magnetic field,
opposite to the diamagnetic Meissner response of the bulk.

In order to perform the experiment proposed in Ref.\ \cite{Igor}, it
is necessary to know the orientation of vector {\bf d}.  The
theoretical analysis \cite{Lebed00} of the $H_{c2}$ anisotropy
\cite{Lee97} indicates that ${\bf d}\|{\bf a}$.  However, recent
Knight shift measurements \cite{private} indicate that ${\bf d}\|{\bf
c}$.  We will consider the former case below.  In the latter case,
magnetic field ${\bf H}\|{\bf c}$ would quickly suppress
superconductivity by the orbital effect.  In the case ${\bf d}\|{\bf
a}$, local magnetic susceptibility could be measures using a coil, as
shown in Fig.\ \ref{fig:coil}, or by another method.  As the coil
approaches to the sample end, magnetic susceptibility should change
sign from diamagnetic to paramagnetic because of the edge states
contribution.  They are localized within the coherence length
$\xi=\hbar v_F/\Delta_0=0.6$ $\mu$m, where $\Delta_0\approx0.22$ meV
is the superconducting gap \cite{TMTSF}.

\subsection{Schottky anomaly in specific heat}

For ${\bf H}\|{\bf d}$, the Zeeman-split edge states effectively form a
two-level system.  It should exhibit the Schottky anomaly in specific
heat:
\begin{equation}
  C_e=\mathcal{N}_c k_B 
  \left(\frac{\mu_BH/2k_BT}{\cosh(\mu_BH/2k_BT)}\right)^2,
  \label{Schottky}
\end{equation}
where $\mathcal{N}_c\approx4\times10^{10}$ is the total number of
chains.  Eq.\ (\ref{Schottky}) reaches the maximum $C_{\rm
max}/\mathcal{N}_c=0.44~k_B=6.1\times 10^{-24}~\rm J/K$ at the
temperature $T_{\rm max}=\mu_BH/2.4k_B$ proportional to the magnetic
field.  In order to avoid the contribution from the extended bulk
states above the gap, the temperature should be lower than the energy
gap: $T\ll\Delta_0$, thus the magnetic field should be well below the
Pauli limiting field: $\mu_BH\ll\Delta_0$.

%%%%%%%%%%%%%%%%%%%%%%%%%%%%%%%%%%%%%%%%%%%%%%%%%%%%%%%%%%%%%%%%%%%%%
\section{Chiral Andreev edge states in the superconducting $\rm Sr_2RuO_4$}
\label{sec:SrRu}
%%%%%%%%%%%%%%%%%%%%%%%%%%%%%%%%%%%%%%%%%%%%%%%%%%%%%%%%%%%%%%%%%%%%%

$\rm Sr_2RuO_4$ is a quasi-two-dimensional perovskite with the
superconducting transition temperature $T_c=1.5$ K. The main Fermi
surface is a cylinder of radius $k_F=7.5\times10^9~\rm m^{-1}$, and
the interlayer spacing is $c=1.3$ nm \cite{Akima}.  We assume the
typical sample dimensions to be $L_x=L_y=2$ mm and $L_z=1.3$ mm, which
makes $\mathcal{N}_l=L_z/c=10^6$ layers.

The superconducting pairing in $\rm Sr_2RuO_4$ is believed to be
triplet and chiral.  The simplest proposed pairing potential has the
form $\Delta({\bf k})=\Delta_0(k_x \pm ik_y)/k_F$ \cite{Rice95}.  More
recently, it was suggested that the gap $\Delta_0$ is a real function
of {\bf k} with the nodes.  The theoretical fit \cite{Krish} of the
tunneling data supports the horizontal lines of nodes, with $\Delta_0$
being a periodic function of $k_z$.  However, in the present paper, we
only consider the simplest case $\Delta_0=\rm const$ without nodes,
because we focus on the question whether the superconductivity in $\rm
Sr_2RuO_4$ is chiral.  The main experimental evidence for that is the
change of muon spin relaxation time at $T_c$ \cite{Luke98}.  However,
this is rather indirect indication of the time-reversal symmetry
breaking.  Below we discuss experiments with the chiral edge states,
which could give direct proofs of the time-reversal symmetry breaking.
These experiments do not depend qualitatively on whether $\Delta_0$ is
constant or modulated.

\subsection{Time-of-flight experiment}
\label{pulse}

A $k_x\pm ik_y$-wave superconductor has chiral Andreev edge states,
which circulate around the perimeter of the sample with the group
velocity $v_e=\Delta_0/\hbar k_F\approx45$ m/s \cite{Honerkamp98},
where we used the value $\Delta_0=2.6$ K.  These states are analogous
to those in QHE systems, e.g.\ the FISDW system discussed in Sec.\
\ref{sec:FISDW}.  The conventional QHE with electric voltage is not
possible in superconductors \cite{Furusaki}, but the spin QHE is
possible \cite{Senthil}.  The chiral character of the edge states in
$\rm Sr_2RuO_4$ could be detected in the time-of-flight experiments
described in Sec.\ \ref{pulse1}, performed with magnetic or thermal,
but not electric pulses.  Suppose a short pulse of a magnetic field
parallel to ${\bf d}$ is applied at a certain point on the edge of the
sample.  The pulse would create a local population imbalance between
the up and down spin states and, thus, a local magnetization.  This
spin imbalance will then travel along the edge with the group velocity
$v_e$, and the corresponding magnetization can be detected at a
distance $L_x$ at the time $t=L_x/v_e$ with a high-sensitivity SQUID
magnetometer \cite{SQUIDs}.  Given the typical sample size, the
time delay could be about $t\approx40$ $\mu$s. The duration of the
pulse should be shorter than $t$, but longer than
$\hbar/\Delta_0\approx3$ ps.  The maximum possible spin imbalance is
achieved when $\mu_BH=\Delta_0$ at $H\approx3.9$ T, which generates
magnetic moment $M_e=\mu_Bk_F/2\pi\approx1.1\times10^{-14}$ A~m per
unit length of the edge.  Note that $\bf H\perp d$ will not produce
the effect.

However, as discussed in Sec.\ \ref{ThermalQH2}, the longitudinal
thermal conductance in $\rm Sr_2RuO_4$ is much greater than the
transverse one, so the edge pulse could quickly diffuse into the bulk.

\subsection{Thermal quantum Hall effect}
\label{ThermalQH2}

The chiral edge states in a $k_x\pm ik_y$-wave superconductor should
produce the thermal QHE \cite{Senthil}.  The magnitude of the thermal
Hall conductance is given by Eq.\ (\ref{Hall}) with $N=1$, and the
factor 2 removed for lack of spin degeneracy.  However, the
longitudinal thermal conductivity in $\rm Sr_2RuO_4$ at $T=0.32$ K is
$\kappa_{xx}/T=4$ W/K$^2$~m \cite{Tanatar00}. This translates into the
thermal conductance $K_{xx}/T=\kappa_{xx}c/T=5.28\times10^{-9}~\rm
W/K^2$ per layer, which is three orders of magnitude larger than the
quantum of thermal Hall conductance.  Thus, the thermal transport in
$\rm Sr_2RuO_4$ should be predominantly longitudinal, and it would be
problematic to detect a very small transverse temperature difference
$\delta_yT$ in the experimental setup shown in Fig.\ \ref{fig:TQHE}.

\subsection{Scanning SQUID imaging of spontaneous magnetic fields}
\label{sec:SQUID}

\begin{figure}[t]
\centerline{\psfig{file=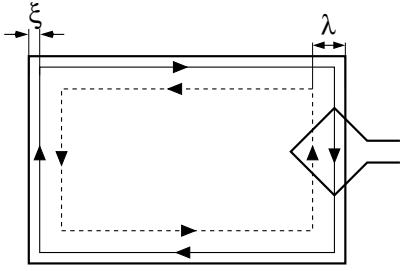,width=0.7\linewidth}}
\caption{Electric currents (solid lines) of the chiral edge states
along the inner and outer edges of $\rm Sr_2RuO_4$.  The dashed lines
show superfluid counterflow.  The rectangular coil on the right
represents a SQUID pickup loop.}
\label{fig:SQUID}
\end{figure}

The chiral edge states in a $k_x\pm ik_y$-wave superconductor also
carry the ground-state electric current $I_e=ek_F^2\hbar/4\pi m=5.6\times
10^{-6}$ A per layer \cite{Matsumoto}, where $m=14.6\,m_e = 1.33\times
10^{-29} $ kg is the effective electron mass.  Multiplied by the
number of layers $\mathcal{N}_l$, this translates into the total
surface current of 5.6 A, which would generate a huge spontaneous
magnetic field in the bulk of the superconductor.  However, this
magnetic field is actually screened by the Meissner supercurrent of
the condensate \cite{Matsumoto,Mineev}.  The distribution of electric
current and magnetic field near the surface of $\rm Sr_2RuO_4$ was
calculated self-consistently in Ref.\ \cite{Matsumoto}, assuming an
infinitely long sample in the {\bf c} direction.  They found that the
edge states current is localized within the coherence length $\xi=66$
nm, whereas the condensate counter-current flows within the larger
penetration depth $\lambda=180$ nm, as sketched in Fig.\
\ref{fig:SQUID}.  Because of the difference between $\xi$ and
$\lambda$, there is a non-zero magnetic field near the surface, with
the maximal value $H_{\rm
max}=0.03\,\Phi_0/2\sqrt{2}\pi\xi\lambda\approx5.88$ G reached at the
distance $\lambda$ and exponentially decreasing inside the bulk. (Here
$\Phi_0=h/2e=2.07\times10^{-15}$ T~m$^2$ is the magnetic flux
quantum.)  This magnetic field produces the magnetic flux
$8.2\times10^{-2}\,\Phi_0/\pi^2\xi=2.6\times10^{-10}$ T~m per unit
length of the edge, which could be detected by a scanning SQUID
microscope \cite{SQUIDs}, as shown in Fig.\ \ref{fig:SQUID}.
The total magnetic flux through the SQUID pickup loop of the size
$L\approx10$ $\mu$m is
$\Phi=8.2\times10^{-2}\Phi_0L/\pi^2\xi\approx1.2\,\Phi_0$, big enough
for SQUID detection.  On the other hand, SQUID microscope would not
resolve the boundaries between domains with opposite chiralities,
where the average magnetic flux is zero \cite{Matsumoto}.

\section{Conclusions}

Many of the experiments discussed above are technically very
challenging.  The most realistic experiment appears to be the
observation of spontaneous magnetic flux at the edges of $\rm
Sr_2RuO_4$ discussed in Sec.\ \ref{sec:SQUID}.  It only requires
cooling the sample below 1 K and scanning the edges with a SQUID
microscope having the typical magnetic flux sensitivity and pickup
loop size \cite{SQUIDs}.  Positive or negative result of such an
experiment would permit to make a definite conclusion whether
superconductivity in $\rm Sr_2RuO_4$ is chiral or not, i.e.\ whether
it breaks time-reversal symmetry.

\end{document}